# Multiqubit entanglement and quantum phase gates with epsilon-near-zero plasmonic waveguides

Ying Li,[1] and Christos Argyropoulos[2,a)]

[1]*School of Physics and Optoelectronic Engineering, Nanjing University of Information Science and Technology, Nanjing, 210044, China*

[2]*Department of Electrical and Computer Engineering, University of Nebraska-Lincoln, Lincoln, NE, 68588, USA*

Multiqubit entanglement is extremely important to perform truly secure quantum optical communication and computing operations. However, the efficient generation of long-range entanglement over extended time periods between multiple qubits randomly distributed in a photonic system remains an outstanding challenge. This constraint is mainly due to the detrimental effects of decoherence and dephasing. To alleviate this issue, we present engineered epsilon-near-zero (ENZ) nanostructures that can maximize the coherence of light-matter interactions at room temperature. We investigate a practical ENZ plasmonic waveguide system which simultaneously achieves multiqubit entanglement in elongated distances, extended time periods, and, even more importantly, independent of the emitters' positions. More specifically, we present efficient transient entanglement between three and four optical qubits mediated by ENZ with results that can be easily generalized to an arbitrary number of emitters. The entanglement between multiple qubits is characterized by computing the negativity metric applied for the first time to the proposed nanophotonic ENZ configuration. The ENZ response is found to be substantially advantageous to boost the coherence between multiple emitters compared to alternative plasmonic waveguide schemes. Finally, the superradiance collective emission response at the ENZ resonance is utilized to design a new high fidelity two-qubit quantum phase gate that can be used in various emerging quantum computing applications.

Quantum entanglement[1] lies at the heart of quantum teleportation, quantum cryptography, and other quantum information processing protocols that are expected to have paramount importance in secure quantum optical communications. However, it is usually achieved at extremely short distances and under very short time periods mainly due to the weak and incoherent interactions between qubits.[2] To overcome this limitation, the coupling of various quantum emitters (a.k.a. qubits) to dissipative plasmonic reservoirs, such as plasmonic waveguides, nanospheres, and nanoantennas, have been proposed,[3,4] resulting in significantly enhanced and coherent dipole-dipole interactions. These plasmonic quantum electrodynamic systems can serve as efficient platforms to achieve effective and long-lived qubit-qubit entanglement.[5] Nevertheless, the vast majority of relevant quantum entanglement studies are focused on the most usual case where the plasmonic system interacts with only two quantum

---

a) Author to whom correspondence should be addressed. Electronic mail: christos.argyropoulos@unl.edu.

emitters.[6–12] Interestingly, only a few studies[13–15] have been focused on the entanglement of multiple emitters coupled to simplified plasmonic systems based on nanoparticles or arrays of them. The metric of concurrence, which is typically used to characterize the entanglement between two qubits, is not applicable to three or more qubits and more complicated quantum optical metrics, such as negativity[16] and genuine multipartite entanglement,[17] should be employed. In addition, the emitters' spatial positions along simplified plasmonic systems (nanoparticles) are crucial to the efficient quantum entanglement performance.[18] In plasmonic nanoparticles or waveguides sustaining surface plasmons, each qubit has to be accurately placed at a specific location where the density of optical states is locally maximized.[19,20] Currently, however, large field enhancement is achieved in plasmonic systems only over very limited spatial extents. In addition, the development of the envisioned integrated quantum photonic circuitry will require efficient coupling between quantum emitters and nanophotonic structures, where, ideally, the coupling will need to be independent of the emitters' spatial locations. Addressing these grand challenges will be the key to enable efficient long-range multiqubit entanglement.

To overcome these limitations, metamaterials exhibiting epsilon-near-zero (ENZ) permittivity response have attracted increased attention[21–24] due to their unique and fascinating features to enable uniform field enhancement in elongated regions leading to numerous potential applications in boosting optical nonlinearities,[25–29] new integrated photonic devices,[30,31] and efficient optical interconnects.[32] Here we investigate an array of engineered plasmonic waveguides that exhibit an effective ENZ response at their cutoff frequency.[25] Strong coupling of the incident light inside each waveguide is achieved, accompanied by a large field enhancement and uniform phase distribution,[33] a combination of properties ideally suited to boost the coherent interaction between different emitters embedded in the nanochannels.[34]

Recently, a variety of quantum optical effects such as enhanced spontaneous emission,[35] boosted superradiance,[36,37] resonance energy transfer and bipartite entanglement[12] have been achieved by using such ENZ plasmonic configurations. In this Letter, we present efficient multipartite entanglement between three and four qubits mediated by ENZ plasmonic waveguides with results that can be easily generalized to an arbitrary emitter number. The entanglement of multiple emitters coupled to other plasmonic waveguide systems is also explored and compared to ENZ. It is explicitly proven that only the ENZ system can substantially generate prolonged entanglement among multiple quantum emitters independent of their positions in the nanowaveguide. Moreover, it is theoretically predicted that the ENZ plasmonic waveguide can realize a deterministic quantum phase gate between two qubits with high fidelity and long separation distances due to the superradiance collective emission state at the ENZ resonance. A novel way to alleviate the inherent losses of metals, and, as a result, reduce decoherence and dephasing, based on the introduction of gain in the nanowaveguides that subsequently increases the fidelity of the proposed quantum phase gate is also presented.

The proposed plasmonic waveguide design is shown in Fig. 1(a), where a narrow rectangular slit of width $w$, height $t<<w$, and length $l$ is carved inside a silver (Ag) screen and loaded with a homogeneous dielectric medium with relative permittivity $\varepsilon = 2.2$. This configuration consists the unit cell of an array of waveguides with periodicities equal to $a$ and $b$. The waveguide's width $w$ is appropriately designed to tune the cutoff frequency of its dominant quasi-transverse electric (quasi-TE) mode to be equal to $295\,\text{THz}$, where an effective ENZ response can be achieved.[25,38] We consider three identical z-oriented quantum emitters, such as quantum dots (QDs), located inside the plasmonic waveguide channel with equal separation distances $d$ to obtain a fair comparison with the response of other plasmonic waveguides, as presented later. Each emitter can be regarded as a qubit with two energy states: ground $|g\rangle$ and excited $|e\rangle$, and we assume that only one qubit is excited initially. We assume that the emitters are sufficiently far from each other, so that interatomic Coulomb interactions can be ignored. All the states of interest and transitions between them are illustrated in Fig. 1(b), where $|U_i\rangle$ represents the excited state of the $i$-th emitter when all the other emitters are in the ground state, and $|ggg\rangle$ is the state where all emitters are in the ground state. For simplicity, the three identical qubits are assumed to have the same transition frequencies, chosen to be equal to the ENZ cutoff frequency. Under the Born-Markov and rotating wave approximations and assuming weak excitation approximation at the weak coupling regime, a master equation for the three-qubit system, represented by the reduced density matrix $\rho(t)$, can be derived and solved numerically (more details provided in the Supplementary[38]).

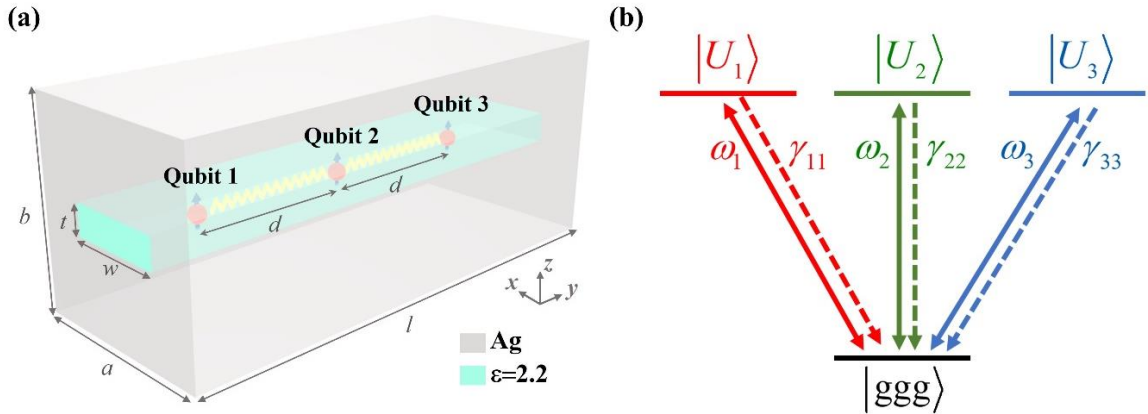

FIG. 1. (a) Geometry of the ENZ plasmonic waveguide with z-oriented emitters embedded in the nanochannel. (b) Quantum states diagram of three emitters.

Once the density matrix $\rho(t)$ is computed, the entanglement of three or more qubits cannot be characterized by the conventional concurrence function,[12] but needs to be quantified by means of alternative more complicated quantum optical metrics, such as the entanglement entropy,[39] state negativity,[15] or genuine multipartite entanglement measurements.[13] More specifically, for a quantum system consisting of two qubits $A$ and $B$, the negativity of a state, which is based on the Peres-Horodecki criterion[40] for entanglement, is defined as:[41]

$$N_{A-B}(\rho) = \max\left(0, -2\sigma_{neg}(\rho^{TA})\right), \tag{1}$$

where $\sigma_{neg}(\rho^{TA})$ is the sum of the negative eigenvalues of the partial transpose $(\rho^{TA})$ of the density matrix $(\rho)$ with respect to qubit $A$.[42] For $N$-qubit quantum systems, the negativity metric can be generalized and defined as $N_{A_1 A_2 \ldots A_n}(\rho) = \left(N_{A_1-A_2\ldots A_n} N_{A_2-A_1 A_3 \ldots A_n} \ldots N_{A_n-A_1 \ldots A_{n-1}}\right)^{1/n}$, where each $A_i$ denotes a qubit. Negativity greater than zero is a sufficient inseparable (entanglement) condition for multipartite quantum systems, indicating that each qubit is inseparable (entangled) from each other.[15] Moreover, we also compute the genuine multipartite entanglement as an additional metric to further quantify the multipartite entangled state with results shown in Supplementary.[38] Note that both metrics lead to similar results.

Hence, we calculate the transient negativity of the ENZ plasmonic waveguide system when using three qubits and compare the results to the simple free space and other commonly used plasmonic waveguide configurations, such as finite groove and rod.[8] The positions of the three emitters in each waveguide system are illustrated in the right insets of Fig. 2, where they are oriented vertically in the ENZ and rod waveguides and horizontally along the groove to maximize their coupling with the ENZ and surface-plasmon mode, respectively. The dipole moments of all emitters are identical, since similar qubits (such as QDs) are expected to be loaded in the plasmonic waveguides. The emission frequency of each emitter corresponds to the ENZ cutoff $(295\,\text{THz})$. In addition, the normalized electric field patterns when one emitter is placed inside or along each waveguide are shown in the right insets of Fig. 2, where it is proven that only the ENZ waveguide has a homogeneous enhanced field distribution. The groove parameters are $L = 235\,\text{nm}$ and $\theta = 10°$, while the rod has radius $R = 25\,\text{nm}$. The waveguide lengths are chosen to be $1\,\mu\text{m}$ for both rod and ENZ cases and $1.4\,\mu\text{m}$ for groove waveguide to ensure the distance between two field antinodes at the groove's end is also $1\,\mu\text{m}$. We study two separation distances for each adjacent emitters: $d = 200\,\text{nm}$ and $d = 400\,\text{nm}$, and consider all potential excitation situations in each qubit, which are $Q_1$ (or $Q_3$ due to symmetry) initially excited, i.e., $\rho_{11}(0) = 1$, and $Q_2$ initially excited, i.e., $\rho_{22}(0) = 1$.

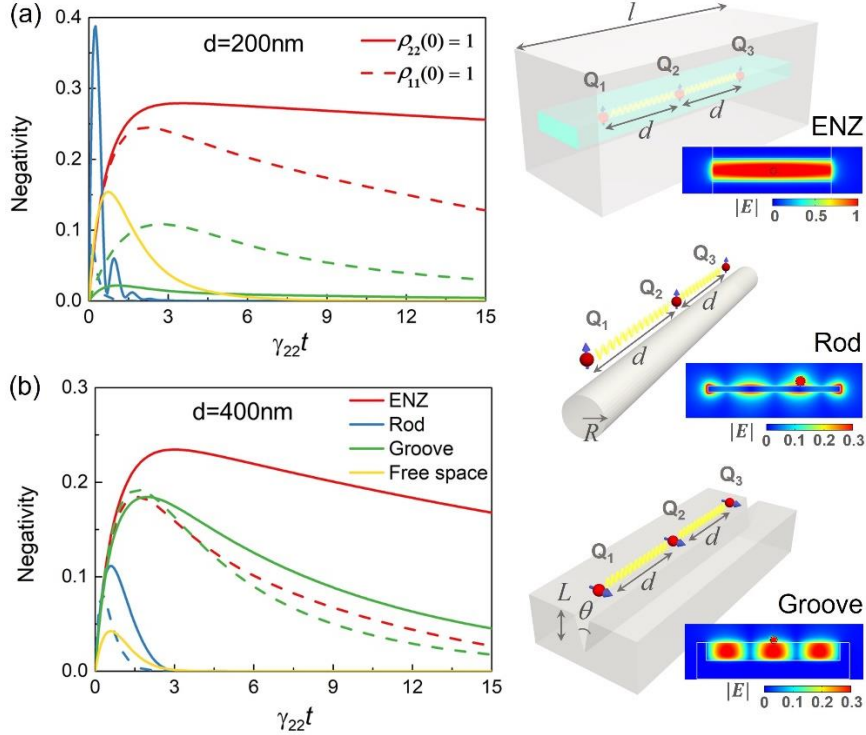

FIG. 2. Negativity for three qubits placed in the ENZ, cylindrical rod, and groove plasmonic waveguides or in free space. The inter-qubit separation distances are (a) $d = 200 nm$ and (b) $d = 400 nm$. The initially excited qubit is $Q_2$ with $\rho_{22}(0) = 1$ or $Q_1$ with $\rho_{11}(0) = 1$. The x-axes are normalized to $\gamma_{22}$, i.e., the computed emission decay rate of $Q_2$ in each scenario. Right insets: plasmonic waveguide schematics and normalized electric field patterns when an emitter is placed inside or along each waveguide.

As shown in Fig. 2, a transient entangled state (corresponding to negativity larger than zero values) can be created in all cases but decays under different timescales. The negativity values are dependent on the initial state of each qubit, e.g., the negativity under $Q_2$ initially excited is always larger than when $Q_1$ is initially excited in the same waveguide system. Note that the transient negativity mediated by the ENZ plasmonic waveguide is superior to other plasmonic waveguides and free space, even for relatively large separation distance $(d = 400 nm)$. This is mainly due to the enhanced homogeneous field mode (Fig. 2 upper right inset and Supplementary Fig. S1(b)) at the cutoff frequency that spreads across the entire ENZ plasmonic waveguide geometry, resulting in constantly large absolute values in the mutual interaction decay rates $\gamma_{ij}$ $(i, j = 1, 2, 3)$ and much lower dipole-dipole interaction rates $g_{ij}$,[38] which are ideal conditions to achieve enhanced entanglement.[12] Interestingly, the tripartite entanglement mediated by the groove waveguide for $d = 200$nm (green lines in Fig. 2(a)) has much lower negativity values than the other waveguide cases. The poor performance of the groove waveguide is somewhat surprising, since a strong surface-plasmon mode is excited along its length (Fig. 2 lower right inset). This problem is primarily due to the standing wave field distribution of the surface-plasmon mode that results in nonuniform decay rate enhancement along its surface.[20] Specifically, $Q_2$ is located at the center (antinode area) of the groove when $d = 200$nm, achieving a rather large decay rate $\gamma_{22} \sim 104\,\text{GHz}$, while $Q_1$ and $Q_3$ are located around the field nodes leading to relative small decay rates $\gamma_{11} = \gamma_{33} \sim 13\,\text{GHz}$. Hence, three qubits cannot be simultaneously entangled to each other due to $\gamma_{11} = \gamma_{33} \ll \gamma_{22}$, naturally leading to small tripartite entanglement. In contrast, the negativity mediated by the rod waveguide for $d = 200\,\text{nm}$ (blue lines in Fig. 2(a)) can rapidly reach very high values accompanied by strong oscillations that are fully dissipated after a very short time duration. This situation usually happens in photonic-crystal or microcavity mediated entanglement[43,44] due to the large dipole-dipole interactions $\left(g_{ij} \gg \gamma_{ii}\right)$. As an example, in the rod waveguide case (Fig. 2 middle right inset), the $Q_2$ is fixed at the center (node region) and has low decay rate $\gamma_{22} \sim 5\,\text{GHz}$, much smaller than the dipole-dipole interaction rate $g_{13} \sim 35\,\text{GHz}$. Note that the entanglement is always studied for qubits placed inside one nanochannel of the ENZ waveguide system, but similar results will be obtained if the qubits are placed in each periodic nanochannel, since the structure's periodicity will not affect its ENZ operation.[25]

Therefore, commonly used plasmonic waveguide configurations, such as groove and rod, achieve low three-qubit entanglement that strongly depends on the spatial position of each emitter, obviously a severe disadvantage for their practical application in quantum technologies since it is extremely difficult to accurately position emitters in nanoscale areas. However, the proposed ENZ plasmonic waveguide can generate efficient and prolonged multiqubit entanglement without being affected by the emitters' separation distance due to the uniform filed amplitude along the entire nanochannel. Hence, the emitters can be randomly distributed inside the ENZ nanochannel, which is a much more practical scenario that will help towards the experimental verification of the concept.

The presented theoretical calculations to quantify multiqubit entanglement can also be extended to four-qubit or even $N$-qubit scenarios. A relevant schematic is shown in Fig. 3(a), where we consider four identical quantum emitters $Q_i$ mediated by a plasmonic waveguide dissipative reservoir. The state transitions are depicted in Fig. 3(b), where all emitters operate at the ENZ cutoff frequency. The more complicated master equation of the four-qubit system is derived in Supplementary.[38] Once the density matrix $\rho(t)$ is computed, the quadripartite entanglement characterized by negativity can be calculated by generalizing the metric of Eq. (1) to four qubits. Assuming that $Q_2$ is initially excited and each inter-qubit separation distance is $d = 300\,\text{nm}$, we plot in Fig. 3(c) the transient negativity for ENZ, finite groove, and rod plasmonic waveguides along with free space. The negativity values are again superior for an extended time duration only with the ENZ system, meaning that an enhanced quadripartite entanglement is realized. Again, this interesting effect is mainly due to the homogeneous field enhancement along the entire nanochannel at the ENZ resonance. On the contrary, the finite groove and rod waveguides exhibit nonuniform standing wave field distributions, similar to a Fabry-Perot cavity modes. As a result, four qubits are less likely to be efficiently entangled to each other, therefore, the associated negativity is lower and decays faster than the ENZ system. Note that the current results can be extended to even larger qubit numbers, subject that they can fit along each waveguide, by generalizing the negative metric given by Eq. (1).

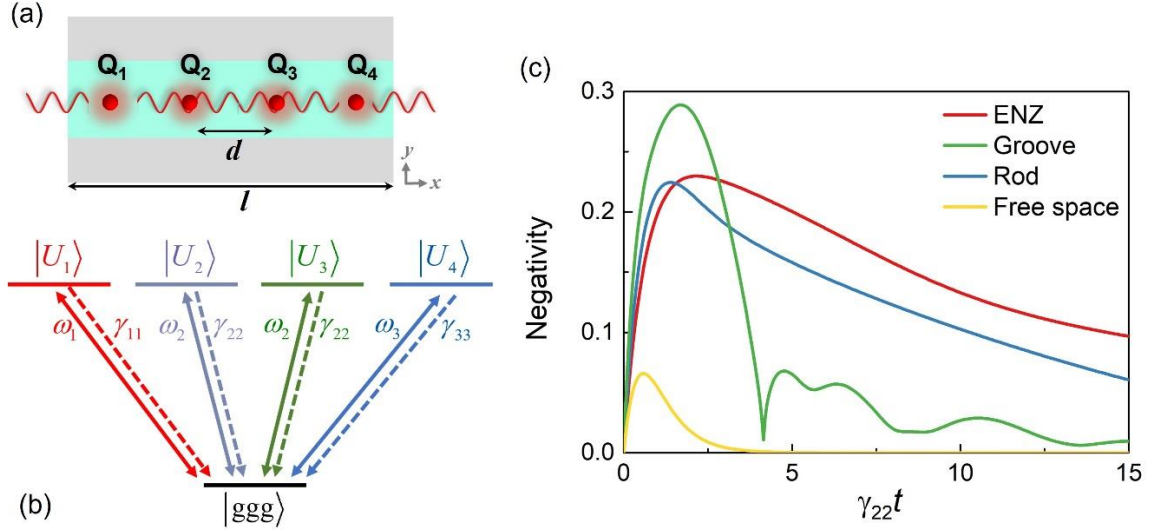

FIG. 3. (a) Schematic of the ENZ plasmonic waveguide loaded with four qubits with similar distance. (b) The corresponding quantum states diagram. (c) Negativity of four qubits placed in ENZ, groove, and rod plasmonic waveguides, or free space. The x-axis is normalized to $\gamma_{22}$, i.e., the computed emission decay rate of $Q_2$ in each scenario.

Finally, we present the realization of a new two-qubit quantum phase gate based on the ENZ plasmonic waveguide that exhibits high fidelity. This quantum phase gate is expected to work by employing the large decay difference between superradiant and subradiant states[18] that exists in the presented ENZ nanowaveguide system.[12,37] As shown in Fig. 4(a), we consider two three-level $\Lambda$-type atoms embedded inside the ENZ waveguide and assume that only the $|e\rangle \to |g\rangle$ transition can be coupled to the ENZ mode through external pumping with strength equal to the effective Rabi frequencies $\Omega_1$ and $\Omega_2$. $|s\rangle$ is a non-resonance auxiliary level of the $\Lambda$-type emitter that does not couple to the ENZ mode. The schematics of the collective states and associated energy levels are given in Fig. 4(b), where $|+\rangle = (|ge\rangle + |eg\rangle)/\sqrt{2}$ and $|-\rangle = (|ge\rangle - |eg\rangle)/\sqrt{2}$ are symmetric and antisymmetric states with collective decay rates $\gamma_+ = \gamma_{11} + \gamma_{12}$ and $\gamma_- = \gamma_{11} - \gamma_{12}$, respectively. It was recently shown[37] that the ENZ waveguide at the cutoff frequency has $\gamma_{11} \sim \gamma_{12}$ along the entire channel, meaning that the symmetric state $|+\rangle$ is superradiant, whereas the antisymmetric state $|-\rangle$ is subradiant $(\gamma_+ \gg \gamma_-)$. Note that a pure superradiant emission and a near-zero subradiant decay rate, independent of the qubits' positions, can be sustained only by the presented ENZ waveguide.[37]

To design an efficient quantum phase gate, we need to introduce a $\pi$ phase shift on the resulted collective ground state $|gg\rangle$.[18,45,46] In general, a $2\pi$ Rabi pulse $(\Omega t = 2\pi)$ will lead to a total interaction time of $t = 2\pi/\Omega$, which will force the system to undergo a full Rabi oscillation, leading to a $\pi$ phase shift in the atom-field state.[47,48] However, if the decay rate of an excited state is much stronger than the Rabi frequency, the driving field cannot result in Rabi oscillations anymore, and instead, introduces scattering.[42] Specifically, the effective driving strengths for $|+\rangle$ and $|-\rangle$ collective states can be written as: $\Omega_+ = (\Omega_1 + \Omega_2)/\sqrt{2}$ and $\Omega_- = (\Omega_1 - \Omega_2)/\sqrt{2}$ for the two-atom system that is presented here. If we choose atoms with Rabi frequencies $\Omega_1 = -\Omega_2$, then $\Omega_+ = 0$ and $\Omega_- = \sqrt{2}\Omega_1$. After selecting $\Omega_-$ to satisfy $\gamma_+ \gg \Omega_- \gg \gamma_-$, which is practical in our ENZ system since $\gamma_+ \gg \gamma_-$, the transition from $|gg\rangle$ to $|+\rangle$ and then $|ee\rangle$ will be blocked due to the strong decay $\gamma_+$, whereas, a $2\pi$ Rabi oscillation will be performed between the allowed transition $|gg\rangle \to |-\rangle$ that results in a $\pi$ phase flip on the state $|gg\rangle$. Furthermore, since $|s\rangle$ is a non-resonance auxiliary state, if we start from $|gs\rangle$ (or $|sg\rangle$) state, only one atom can interact with the waveguide with strong decay rate $\gamma_{11} \sim \gamma_+/2$. The transition $|gs\rangle \to |es\rangle$ (or $|sg\rangle \to |se\rangle$) is also blocked due to the driving signal being too weak to excite the atoms.[18] Thus, if we initially prepare the system in a starting state $|\psi_{initial}\rangle = \frac{1}{2}(|ss\rangle + |sg\rangle + |gs\rangle + |gg\rangle)$, by aid of external classical $2\pi$ laser pulses, the final state will turn to $|\psi_{final}\rangle = \frac{1}{2}(|ss\rangle + |sg\rangle + |gs\rangle - |gg\rangle)$ due to the resulted $\pi$ phase shift of the $|gg\rangle$ state, meaning that a two-qubit quantum phase gate is realized.

The gate efficiency can be measured by the metric of fidelity $F$ given by the relation: $F = 1 - \sqrt{\gamma_-/\gamma_{11}}$.[18] Clearly, our ENZ waveguide, exhibiting a near-zero subradiant decay $\gamma_-$,[37] is expected to lead to extremely high gate fidelity, an essential response to achieve an efficient quantum phase gate that can be used in quantum computing and communication systems. Figure 4(c) demonstrates the calculated fidelity as a function of the inter-emitter separation distance in the passive ENZ plasmonic waveguide and free space. The quantum phase gate with high fidelity in free space can be achieved only when the emitters are very closely packed to each other due to the resulted superradiance phenomenon only in highly subwavelength regions. In the passive ENZ case, where the waveguide shown in the left inset of Fig. 4(c) is loaded with a lossless dielectric material $(\varepsilon = 2.2)$, a maximum fidelity of nearly 100% can be achieved for extremely subwavelength separation distances between emitters that decreases monotonously when this distance is increased. This is mainly due to the radiative and inherent Ohmic losses of the metal waveguide system leading to decoherence and dephasing. To further alleviate this issue, we insert a layer of an active medium with length $l_{ac} = 200$ nm in the ENZ nanochannel through choosing a permittivity with a positive imaginary part ($\varepsilon=2.2+i0.045$) (leading to an exceptional point[49]) that fully compensates the inherent plasmonic losses. This low gain value is practical and can be achieved by Ruthenium (Ru) dyes.[50] Interestingly, the gate fidelity in the active ENZ waveguide design can be further increased compared to the passive ENZ case and is only limited by radiation losses. Note that $d$ is always larger than $l_{ac}$ ($d > l_{ac}$) in Fig. 4(c) meaning that both quantum emitters are always inserted in the lossless passive dielectric material.

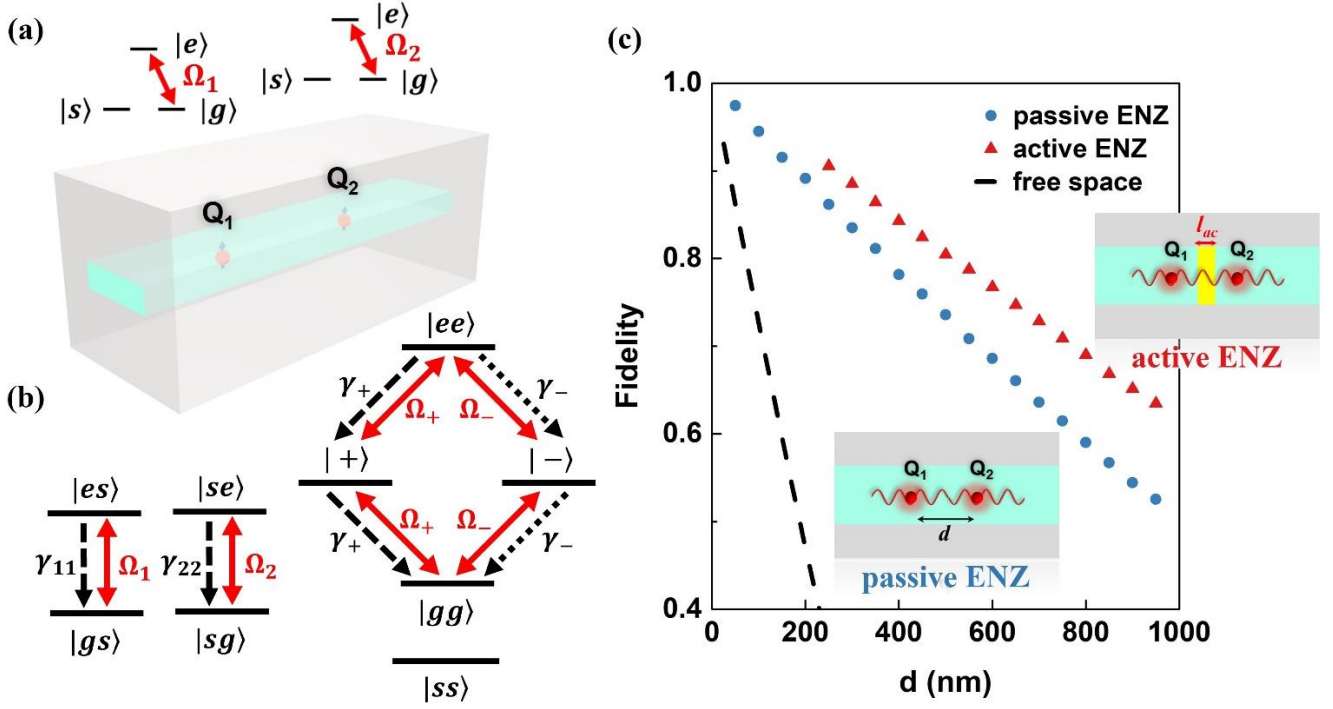

FIG. 4. (a) Two Λ atoms ($Q_1$ and $Q_2$) embedded inside the ENZ nanochannel. The $|e\rangle - |g\rangle$ transitions resonantly couple to the ENZ mode. (b) Diagram of the two Λ atoms energy levels. (c) Gate fidelity as a function of inter-qubit separation distance when passive and active ENZ plasmonic waveguides or free space are used.

In conclusion, a plasmonic waveguide with effective ENZ response has been presented to achieve both transient multiqubit entanglement and two-qubit quantum phase gate functionalities at the nanoscale. Strong and uniform field enhancement is achieved at the ENZ resonance, leading to prolonged and efficient transient entanglement along elongated regions among three and four quantum emitters randomly distributed in this plasmonic waveguide system. These results can be easily generalized to an arbitrary qubit number. Moreover, due to the collective superradiance decay at the ENZ frequency, the plasmonic waveguide can realize an efficient quantum phase gate between two qubits with high fidelity achieved even for long qubit separation distance.

## ACKNOWLEDGMENTS

This work was partially supported by the National Science Foundation (Grant No. DMR-1709612), the Nebraska Materials Research Science and Engineering Center (Grant No. DMR-1420645) and the National Science Foundation/EPSCoR RII Track-1: Emergent Quantum Materials and Technologies (EQUATE) under (Grant No. OIA-2044049). Y. L. is supported by the National Natural Science Foundation of China (Grant No. 12104233) and the Natural Science Foundation of Jiangsu Higher Education Institutions of China (Grant No. 21KJB140013).

## DATA AVAILABILITY

The data that support the findings of this study are available from the corresponding author upon reasonable request.

# *Supplemental Material:* Multiqubit entanglement and quantum phase gates with epsilon-near-zero plasmonic waveguides

Ying Li,[1] and Christos Argyropoulos[2,*]

[1]*School of Physics and Optoelectronic Engineering, Nanjing University of Information Science and Technology, Nanjing, 210044, China*

[2]*Department of Electrical and Computer Engineering, University of Nebraska-Lincoln, Lincoln, NE, 68588, USA*

[*]christos.argyropoulos@unl.edu

**I. Effective epsilon-near-zero (ENZ) resonance response of plasmonic waveguides**

In this work, the proposed plasmonic waveguide design is shown in Fig. 1(a) in the main paper, where a narrow rectangular slit of width *w*, height *t<<w*, and length *l* is carved inside a silver (Ag) screen and loaded with a homogeneous dielectric medium with relative permittivity $\varepsilon = 2.2$. This design is the unit cell of an array of waveguides with periodicities equal to *a* and *b*. The waveguide width is designed to operate at the cutoff of its dominant quasi-transverse electric (quasi-TE) mode, i.e., at the cutoff frequency for which its guided wave number $\beta_{wg}$ has a near-zero real part.[1] Hence, the guided wavelength in each nanowaveguide is effectively infinite, since $\lambda g=2\pi/Re[\beta_{wg}]$. Figure S1(a) shows the calculated effective permittivity $\varepsilon_{eff}$ and dispersion of the guided wave number $\beta_{wg}$ of the system by using the same dimensions used in the main paper: *w* = 200 nm, *t* = 40 nm, *l* = 1 μm, *a* =*b* = 400 nm. At the cutoff frequency (295THz), both the effective permittivity $\varepsilon_{eff}$ (red line) and guided wave number $\beta_{wg}$ (blue line) have a near-zero real part, meaning that an effective ENZ response can be achieved based on this design. Figure S1(b) demonstrates the static electric field pattern and the $E_z$ component phase progression when a quantum dipole emitter with vertically (along *z*-direction) orientation is placed inside the nanochannel operating at the ENZ resonance. The dipole moment of the emitter is chosen to be $\mu = 60D$, typical value for self-assembled quantum dots (QDs),[2] and the ENZ resonance is around the waveguide's cutoff frequency. Note that the mode generated by the *z*-oriented dipole is strongly enhanced in the elongated nanochannel and, even more importantly, almost uniform phase distribution is observed along the entire waveguide length $l = 1$μm. These properties are extremely well suited to boost the coherent interaction between several emitters embedded in the nanochannel.

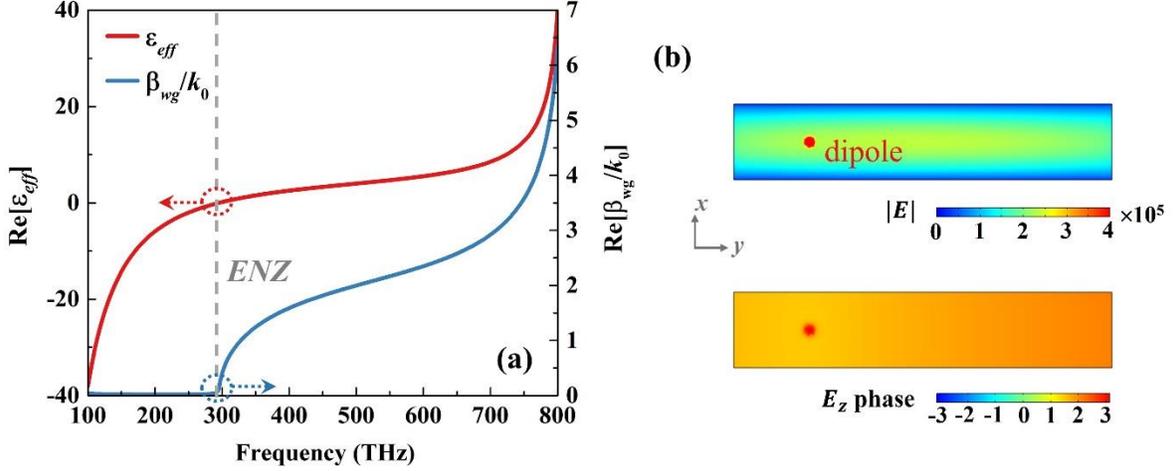

FIG. S1. (a) Real part of the computed effective permittivity (red line, left) and guided wave number (blue line, right) of the plasmonic waveguide versus the incident frequency. (b) Electric field pattern and $E_z$ phase distribution when a dipole emitter oriented along the $z$-direction is placed inside the waveguide.

## II. Master equation for three quantum emitters

Under the Born-Markov and rotating wave approximations and after tracing out the electromagnetic field degrees of freedom, the reduced density matrix $\rho(t)$ for the three emitters system in the weak coupling regime obeys the master equation:[3–5]

$$\frac{\partial \rho}{\partial t} = \frac{1}{2}\sum_{i,j}\gamma_{ij}\left(2\sigma_i \rho \sigma_j^\dagger - \rho \sigma_i^\dagger \sigma_j - \sigma_i^\dagger \sigma_j \rho\right) + \frac{1}{i\hbar}[H,\rho], \qquad (1)$$

where $i,j = 1, 2, 3$ in the three qubit case. The Hamiltonian $H$ is equal to $H = \sum_i \hbar(\omega_0 + g_{ii})\sigma_i^\dagger \sigma_i + \sum_{i\neq j}\hbar g_{ij}\sigma_i^\dagger \sigma_j$ and the two-level emitter is described by the Pauli operator $\hat{\sigma}_i\ (\hat{\sigma}_i^\dagger)$ which is the lowering (raising) operator of the $i$-th qubit. Here, $g_{ii}$ is the single-emitter Lamb shift due to the self-interaction of each qubit and $g_{ij}$ is the radiative coherent dipole-dipole shift. Note that $g_{ii}$ is typically very small, less than few GHz, at optical frequencies for emitter-metal separation distances larger than 10 nm[6] and is neglected in our analysis. Similarly, $\gamma_{ii}$ is the single-emitter spontaneous decay rate and $\gamma_{ij}$ is the noncoherent decay which describes the coupling effect of emitters upon each other. The $g$ and $\gamma$ coefficients can be computed by the Green function of the system under study and given by the formulas:[4,7]

$$g_{ij} = \left(\omega_0^2/\varepsilon_0 \hbar c^2\right)\mathrm{Re}[\boldsymbol{\mu}_i^* \cdot \overline{\mathbf{G}}(\mathbf{r}_i,\mathbf{r}_j,\omega_0)\cdot \boldsymbol{\mu}_j], \qquad (2)$$

$$\gamma_{ij} = \left(2\omega_0^2 / \varepsilon_0 \hbar c^2\right) \text{Im}[\boldsymbol{\mu}_i^* \cdot \overline{\mathbf{G}}\left(\mathbf{r}_i, \mathbf{r}_j, \omega_0\right) \cdot \boldsymbol{\mu}_j], \quad (3)$$

where $\boldsymbol{\mu}_i^*$ is the complex conjugate of the dipole moment of the $i$-th emitter. The dyadic Green function $\overline{\mathbf{G}}\left(\mathbf{r}_i, \mathbf{r}_j, \omega_0\right)$ at position $\mathbf{r}_i$ is the induced field in the system resulted from a delta function excitation at an arbitrary position $\mathbf{r}_j$.[8] It can be defined by the electric field $\mathbf{E}$ at the position $\mathbf{r}_i$ induced by an electric dipole at the source position $\mathbf{r}_j$ using the relationship:

$$\mathbf{E}(\mathbf{r}_i) = \omega_0^2 \mu_0 \mu \overline{\mathbf{G}}\left(\mathbf{r}_i, \mathbf{r}_j, \omega_0\right) \boldsymbol{\mu}, \quad (4)$$

where $\mu_0$ is the permeability of free space and $\mu$ is the relative permeability of the surrounding space. Therefore, in plasmonic systems, the $g$ and $\gamma$ decay rates are directly proportional to the real and imaginary parts of the corresponding mode electric field values, respectively. As shown in the right insets in Fig. 2 in the main paper, the groove and rod surface-plasmon modes have nonuniform standing wave electric field distributions, leading to nonuniform $g$ and $\gamma$ decay rates in terms of the emitter spatial locations. These standing wave distributions are not because of the evanescent nature of the resulted mode profile, but they are actually due to the excited surface-plasmon modes along the finite structures of groove and rod waveguides that have a similar standing-wave field distribution to a Fabry-Perot cavity mode. Finally, it is worth mentioning that $\gamma_{ij} \gg g_{ij}$ is the ideal condition to achieve strong entanglement in the majority of plasmonic-mediated dissipative-driven systems,[4,7,9,10] but is not valid for other waveguide types, such as photonic-crystals or microcavities.[11,12] This is due to the fact that a 90 degrees phase difference between the real and imaginary parts of the system's Green function exist only in the case of plasmonic systems.

Both Eqs. (2) and (3) can be calculated either analytically or numerically through solving classical Maxwell's equations. To simplify the expression, some standard notations are introduced to define the normal basis in the case of three qubits: $|1\rangle \equiv |U_1\rangle$, $|2\rangle \equiv |U_2\rangle$, $|3\rangle \equiv |U_3\rangle$. If we further consider $\gamma_{ij} = \gamma_{ji}$, $g_{ij} = g_{ji}$ (due to reciprocity) and assume that only one qubit is initially excited, the time evolution of the density matrix $\rho(t)$ in Eq. (1) can be expanded into 9 coupled differential equations as follow, which we solve numerically:

$$\dot{\rho}_{11} = -\gamma_{11}\rho_{11} - \frac{\gamma_{12}}{2}(\rho_{21} + \rho_{12}) - \frac{\gamma_{13}}{2}(\rho_{31} + \rho_{13}) + ig_{12}(\rho_{21} - \rho_{12}) + ig_{13}(\rho_{31} - \rho_{13}) \quad (5)$$

$$\dot{\rho}_{12} = -\frac{\gamma_{11}+\gamma_{22}}{2}\rho_{12} - \frac{\gamma_{12}}{2}(\rho_{11}+\rho_{22}) - \frac{\gamma_{13}}{2}\rho_{32} - \frac{\gamma_{23}}{2}\rho_{13} + ig_{12}(\rho_{22}-\rho_{11}) + ig_{13}\rho_{32} - ig_{23}\rho_{13} \quad (6)$$

$$\dot{\rho}_{13} = -\frac{\gamma_{11}+\gamma_{33}}{2}\rho_{13} - \frac{\gamma_{13}}{2}(\rho_{33}+\rho_{11}) - \frac{\gamma_{12}}{2}\rho_{23} - \frac{\gamma_{23}}{2}\rho_{12} + ig_{13}(\rho_{33}-\rho_{11}) + ig_{12}\rho_{23} - ig_{23}\rho_{12} \quad (7)$$

$$\dot{\rho}_{21} = -\frac{\gamma_{11}+\gamma_{22}}{2}\rho_{21} - \frac{\gamma_{12}}{2}(\rho_{11}+\rho_{22}) - \frac{\gamma_{13}}{2}\rho_{23} - \frac{\gamma_{23}}{2}\rho_{31} + ig_{12}(\rho_{11}-\rho_{22}) - ig_{13}\rho_{23} + ig_{23}\rho_{31} \quad (8)$$

$$\dot{\rho}_{22} = -\gamma_{22}\rho_{22} - \frac{\gamma_{12}}{2}(\rho_{21}+\rho_{12}) - \frac{\gamma_{23}}{2}(\rho_{32}+\rho_{23}) + ig_{12}(\rho_{12}-\rho_{21}) + ig_{23}(\rho_{32}-\rho_{23}) \quad (9)$$

$$\dot{\rho}_{23} = -\frac{\gamma_{22}+\gamma_{33}}{2}\rho_{23} - \frac{\gamma_{23}}{2}(\rho_{33}+\rho_{22}) - \frac{\gamma_{12}}{2}\rho_{13} - \frac{\gamma_{13}}{2}\rho_{21} + ig_{23}(\rho_{33}-\rho_{22}) + ig_{12}\rho_{13} - ig_{13}\rho_{21} \quad (10)$$

$$\dot{\rho}_{31} = -\frac{\gamma_{11}+\gamma_{33}}{2}\rho_{31} - \frac{\gamma_{13}}{2}(\rho_{33}+\rho_{11}) - \frac{\gamma_{12}}{2}\rho_{32} - \frac{\gamma_{23}}{2}\rho_{21} + ig_{13}(\rho_{11}-\rho_{33}) - ig_{12}\rho_{32} + ig_{23}\rho_{21} \quad (11)$$

$$\dot{\rho}_{32} = -\frac{\gamma_{22}+\gamma_{33}}{2}\rho_{32} - \frac{\gamma_{23}}{2}(\rho_{33}+\rho_{22}) - \frac{\gamma_{12}}{2}\rho_{31} - \frac{\gamma_{13}}{2}\rho_{12} + ig_{23}(\rho_{22}-\rho_{33}) - ig_{12}\rho_{31} + ig_{13}\rho_{12} \quad (12)$$

$$\dot{\rho}_{33} = -\gamma_{33}\rho_{33} - \frac{\gamma_{23}}{2}(\rho_{32}+\rho_{23}) - \frac{\gamma_{13}}{2}(\rho_{13}+\rho_{31}) + ig_{23}(\rho_{23}-\rho_{32}) + ig_{13}(\rho_{13}-\rho_{31}). \quad (13)$$

### III. Master equation for four quantum emitters

The general master equation for four-qubit system can also be expressed as Eq. (1), where $i, j = 1, 2, 3, 4$ for four qubits and the interaction decay $g_{ij}$ and noncoherent decay $\gamma_{ij}$ are again computed by Eqs. (2)-(3). In the basis of $|1\rangle \equiv |U_1\rangle$, $|2\rangle \equiv |U_2\rangle$, $|3\rangle \equiv |U_3\rangle$, $|4\rangle \equiv |U_4\rangle$ and assuming that only one qubit is initially excited, the time evolution of the density matrix $\rho(t)$ in Eq. (1) can be expanded into 16 coupled differential equations as follows:

$$\dot{\rho}_{11} = -\gamma_{11}\rho_{11} - \frac{\gamma_{12}}{2}(\rho_{21}+\rho_{12}) - \frac{\gamma_{13}}{2}(\rho_{31}+\rho_{13}) - \frac{\gamma_{14}}{2}(\rho_{41}+\rho_{14}) + ig_{12}(\rho_{21}-\rho_{12}) + ig_{13}(\rho_{31}-\rho_{13}) + ig_{14}(\rho_{41}-\rho_{14})$$

$$(14)$$

$$\dot{\rho}_{12} = -\frac{\gamma_{11}+\gamma_{22}}{2}\rho_{12} - \frac{\gamma_{12}}{2}(\rho_{11}+\rho_{22}) - \frac{\gamma_{13}}{2}\rho_{32} - \frac{\gamma_{23}}{2}\rho_{13} - \frac{\gamma_{14}}{2}\rho_{42} - \frac{\gamma_{24}}{2}\rho_{14} + ig_{12}(\rho_{22}-\rho_{11}) + ig_{13}\rho_{32} - ig_{23}\rho_{13} + ig_{14}\rho_{42} - ig_{24}\rho_{14}$$

$$(15)$$

$$\dot{\rho}_{13} = -\frac{\gamma_{11}+\gamma_{33}}{2}\rho_{13} - \frac{\gamma_{13}}{2}(\rho_{33}+\rho_{11}) - \frac{\gamma_{12}}{2}\rho_{23} - \frac{\gamma_{23}}{2}\rho_{12} - \frac{\gamma_{14}}{2}\rho_{43} - \frac{\gamma_{34}}{2}\rho_{14} + ig_{13}(\rho_{33}-\rho_{11}) + ig_{12}\rho_{23} - ig_{23}\rho_{12} + ig_{14}\rho_{43} - ig_{34}\rho_{14}$$

$$(16)$$

$$\dot{\rho}_{14} = -\frac{\gamma_{11}+\gamma_{44}}{2}\rho_{14} - \frac{\gamma_{14}}{2}(\rho_{11}+\rho_{44}) - \frac{\gamma_{12}}{2}\rho_{24} - \frac{\gamma_{13}}{2}\rho_{34} - \frac{\gamma_{24}}{2}\rho_{12} - \frac{\gamma_{34}}{2}\rho_{13} + ig_{14}(\rho_{44}-\rho_{11}) + ig_{12}\rho_{24} - ig_{24}\rho_{12} + ig_{13}\rho_{34} - ig_{34}\rho_{13}$$

(17)

$$\dot{\rho}_{21} = -\frac{\gamma_{11}+\gamma_{22}}{2}\rho_{21} - \frac{\gamma_{12}}{2}(\rho_{11}+\rho_{22}) - \frac{\gamma_{13}}{2}\rho_{23} - \frac{\gamma_{23}}{2}\rho_{31} - \frac{\gamma_{14}}{2}\rho_{24} - \frac{\gamma_{24}}{2}\rho_{41} + ig_{12}(\rho_{11}-\rho_{22}) - ig_{13}\rho_{23} + ig_{23}\rho_{31} - ig_{14}\rho_{24} + ig_{24}\rho_{41}$$

(18)

$$\dot{\rho}_{22} = -\gamma_{22}\rho_{22} - \frac{\gamma_{12}}{2}(\rho_{21}+\rho_{12}) - \frac{\gamma_{23}}{2}(\rho_{32}+\rho_{23}) - \frac{\gamma_{24}}{2}(\rho_{42}+\rho_{24}) + ig_{12}(\rho_{12}-\rho_{21}) + ig_{23}(\rho_{32}-\rho_{23}) + ig_{24}(\rho_{42}-\rho_{24})$$

(19)

$$\dot{\rho}_{23} = -\frac{\gamma_{22}+\gamma_{33}}{2}\rho_{23} - \frac{\gamma_{23}}{2}(\rho_{33}+\rho_{22}) - \frac{\gamma_{12}}{2}\rho_{13} - \frac{\gamma_{13}}{2}\rho_{21} - \frac{\gamma_{24}}{2}\rho_{43} - \frac{\gamma_{34}}{2}\rho_{24} + ig_{23}(\rho_{33}-\rho_{22}) + ig_{12}\rho_{13} - ig_{13}\rho_{21} + ig_{24}\rho_{43} - ig_{34}\rho_{24}$$

(20)

$$\dot{\rho}_{24} = -\frac{\gamma_{22}+\gamma_{44}}{2}\rho_{24} - \frac{\gamma_{24}}{2}(\rho_{22}+\rho_{44}) - \frac{\gamma_{12}}{2}\rho_{14} - \frac{\gamma_{14}}{2}\rho_{21} - \frac{\gamma_{23}}{2}\rho_{34} - \frac{\gamma_{34}}{2}\rho_{23} + ig_{24}(\rho_{44}-\rho_{22}) + ig_{12}\rho_{14} - ig_{14}\rho_{21} + ig_{23}\rho_{34} - ig_{34}\rho_{23}$$

(21)

$$\dot{\rho}_{31} = -\frac{\gamma_{11}+\gamma_{33}}{2}\rho_{31} - \frac{\gamma_{13}}{2}(\rho_{33}+\rho_{11}) - \frac{\gamma_{12}}{2}\rho_{32} - \frac{\gamma_{23}}{2}\rho_{21} - \frac{\gamma_{14}}{2}\rho_{34} - \frac{\gamma_{34}}{2}\rho_{41} + ig_{13}(\rho_{11}-\rho_{33}) - ig_{12}\rho_{32} + ig_{23}\rho_{21} - ig_{14}\rho_{34} + ig_{34}\rho_{41}$$

(22)

$$\dot{\rho}_{32} = -\frac{\gamma_{22}+\gamma_{33}}{2}\rho_{32} - \frac{\gamma_{23}}{2}(\rho_{33}+\rho_{22}) - \frac{\gamma_{12}}{2}\rho_{31} - \frac{\gamma_{13}}{2}\rho_{12} - \frac{\gamma_{24}}{2}\rho_{34} - \frac{\gamma_{34}}{2}\rho_{42} + ig_{23}(\rho_{22}-\rho_{33}) - ig_{12}\rho_{31} + ig_{13}\rho_{12} - ig_{24}\rho_{34} + ig_{34}\rho_{42}$$

(23)

$$\dot{\rho}_{33} = -\gamma_{33}\rho_{33} - \frac{\gamma_{23}}{2}(\rho_{32}+\rho_{23}) - \frac{\gamma_{13}}{2}(\rho_{13}+\rho_{31}) - \frac{\gamma_{34}}{2}(\rho_{43}+\rho_{34}) + ig_{23}(\rho_{23}-\rho_{32}) + ig_{13}(\rho_{13}-\rho_{31}) + ig_{34}(\rho_{43}-\rho_{34})$$

(24)

$$\dot{\rho}_{34} = -\frac{\gamma_{33}+\gamma_{44}}{2}\rho_{34} - \frac{\gamma_{34}}{2}(\rho_{33}+\rho_{44}) - \frac{\gamma_{13}}{2}\rho_{14} - \frac{\gamma_{14}}{2}\rho_{31} - \frac{\gamma_{23}}{2}\rho_{24} - \frac{\gamma_{24}}{2}\rho_{32} + ig_{34}(\rho_{44}-\rho_{33}) - ig_{14}\rho_{31} + ig_{13}\rho_{14} - ig_{24}\rho_{32} + ig_{23}\rho_{24}$$

(25)

$$\dot{\rho}_{41} = -\frac{\gamma_{11}+\gamma_{44}}{2}\rho_{41} - \frac{\gamma_{14}}{2}(\rho_{44}+\rho_{11}) - \frac{\gamma_{12}}{2}\rho_{42} - \frac{\gamma_{24}}{2}\rho_{21} - \frac{\gamma_{13}}{2}\rho_{43} - \frac{\gamma_{34}}{2}\rho_{31} + ig_{14}(\rho_{11}-\rho_{44}) - ig_{12}\rho_{42} + ig_{24}\rho_{21} - ig_{13}\rho_{43} + ig_{34}\rho_{31}$$

(26)

$$\dot{\rho}_{42} = -\frac{\gamma_{22}+\gamma_{44}}{2}\rho_{42} - \frac{\gamma_{24}}{2}(\rho_{22}+\rho_{44}) - \frac{\gamma_{12}}{2}\rho_{41} - \frac{\gamma_{14}}{2}\rho_{12} - \frac{\gamma_{23}}{2}\rho_{43} - \frac{\gamma_{34}}{2}\rho_{32} + ig_{24}(\rho_{22}-\rho_{44}) - ig_{12}\rho_{41} + ig_{14}\rho_{12} - ig_{23}\rho_{43} + ig_{34}\rho_{32}$$

(27)

$$\dot{\rho}_{43} = -\frac{\gamma_{33}+\gamma_{44}}{2}\rho_{43} - \frac{\gamma_{34}}{2}(\rho_{33}+\rho_{44}) - \frac{\gamma_{13}}{2}\rho_{41} - \frac{\gamma_{14}}{2}\rho_{13} - \frac{\gamma_{23}}{2}\rho_{42} - \frac{\gamma_{24}}{2}\rho_{23} + ig_{34}(\rho_{33}-\rho_{44}) - ig_{13}\rho_{41} + ig_{14}\rho_{13} - ig_{23}\rho_{42} + ig_{24}\rho_{23}$$

(28)

$$\dot{\rho}_{44} = -\gamma_{44}\rho_{44} - \frac{\gamma_{14}}{2}(\rho_{41}+\rho_{14}) - \frac{\gamma_{24}}{2}(\rho_{42}+\rho_{24}) - \frac{\gamma_{34}}{2}(\rho_{43}+\rho_{34}) + ig_{14}(\rho_{14}-\rho_{41}) + ig_{24}(\rho_{24}-\rho_{42}) + ig_{34}(\rho_{34}-\rho_{43}) \ .$$

(29)

**IV. Additional metric to compute multipartite entanglement**

In order to identify the multipartite entangled mixed state more precisely and further verify the negativity results presented in the main paper, we also consider the genuine multipartite entanglement $E_G^{(2)}$, a generalization of the global entanglement metric for $N$-qubits,[13] which is defined as:[14]

$$E_G^{(2)} = \frac{2}{N(N-1)}\sum_{l=1}^{N-1}(N-1)G(2,l), \tag{30}$$

where the function $G(2,l) = \frac{4}{3}\left[1 - \frac{1}{N-l}\sum_{j=1}^{N-l}\text{Tr}(\rho_{j,j+l}^2)\right]$ and $\rho_{j,j+l}$ is the reduced density matrix of qubits $j$ and $j+l$ obtained by tracing out the other $N$-2 qubits. The genuine multipartite entanglement $E_G^{(2)}$ is the mean of all $G(2,l)$ values and gives the mean linear entropy of all qubit pairs with the remaining qubits in the system. It can be used to successfully detect different kinds of multipartite entangled states, e.g., the Greenberger-Horne-Zeilinger (GHZ) state $|GHZ\rangle = 1/\sqrt{2}(|ggg\rangle + |eee\rangle)$ and the W state $|W\rangle = 1/\sqrt{3}(|U_1\rangle + |U_2\rangle + |U_3\rangle)$ that have $E_{G\,(GHZ)}^{(2)} = 2/3$ and $E_{G\,(W)}^{(2)} = 16/27$, respectively, in the case of three qubits.[15]

Next, we calculate the transient genuine multipartite entanglement in the ENZ plasmonic waveguide system when using three qubits and compare the results to another two commonly used plasmonic waveguides and free space. The initially excited qubit is also set to be $Q_1$ with $\rho_{11}(0)=1$ or $Q_2$ with $\rho_{22}(0)=1$, similar to the analysis in the main paper. Note that in all our calculations we only consider three identical qubits embedded along or inside the various plasmonic waveguides without the addition of any external pumping mechanism. The initialization state is that one emitter is in the excited state and the other emitters are in the ground state. Our system operates in the weak excitation and coupling regimes and does not include Rabi oscillations (or other stimulated emission processes), which means that the three-qubit dynamics have a purely dissipative nature. Therefore, we only consider the spontaneous decay of a single excitation and explore the situation of spontaneous formation of transient entanglement from an unentangled state, similar to various previous relevant works.[7,9,10,16,17] We clearly observe the tripartite entangled state from the genuine multipartite entanglement $E_G^{(2)}$ results shown in Figs. S2(a)-(b), where the decay trends are consistent with the negativity results in Figs. 2(a)-(b) in the main paper. Note that in the ENZ waveguide case, the maximum $E_G^{(2)}$ value can nearly reach to 16/27 and decay very slow, a value approximately equal to the ideal W tripartite entangled state ($|W\rangle = 1/\sqrt{3}(|U_1\rangle + |U_2\rangle + |U_3\rangle)$). This result means that the ENZ-mediated three-qubit system can achieve a maximally entangled state in the weak-coupling regime.

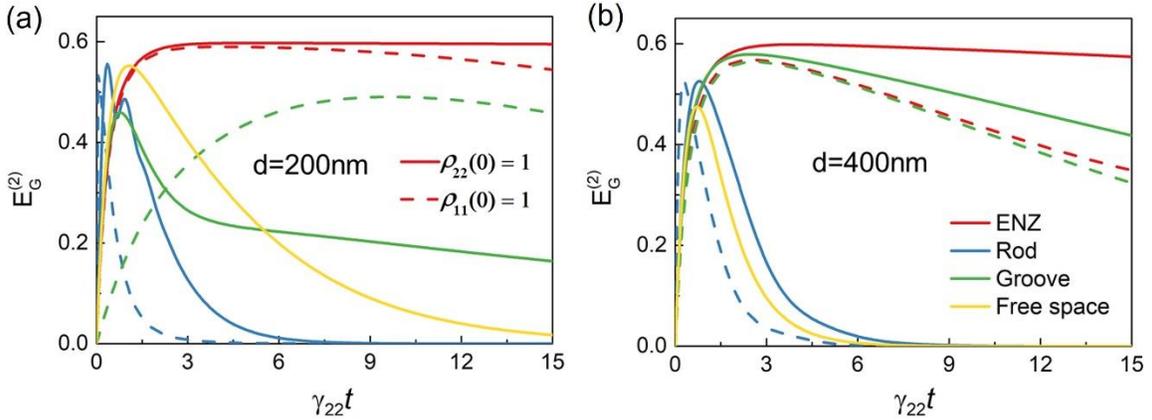

FIG. S2. Genuine multipartite entanglement $E_G^{(2)}$ computed for three qubits placed in the ENZ (red lines), cylindrical rod (blue lines), and groove (green lines) plasmonic waveguides or in free space (yellow lines). The inter-qubits separation distances are (a) $d = 200$nm and (b) $d = 400$nm. The initially excited qubit is set to be $Q_2$ with $\rho_{22}(0) = 1$ (solid lines) and $Q_1$ (or $Q_3$ due to symmetry) with $\rho_{11}(0) = 1$ (dashed lines). The x-axes are normalized to $\gamma_{22}$, i.e., the computed emission decay rate of $Q_2$ in each scenario.

Moreover, we plot the time evolution of the different excited state populations $\rho_{11}$, $\rho_{22}$, and $\rho_{33}$ under the normal basis when the three qubits are embedded inside the ENZ, cylindrical rod, and groove plasmonic waveguide. The results are shown in Fig. S3, where the inter-qubit separation distance is fixed to $d = 200$ nm and the initially excited qubit is assumed to be $Q_2$ with $\rho_{22}(0) = 1$. We can see that the decay of the emitter initially in the excited state ($Q_2$) leads to excitation of the emitters initially in the ground states ($Q_1$ and $Q_3$). As a result, in each waveguide system, a transient entangled state can be created but decays under different timescales. The longest decay time is achieved by the ENZ waveguide, as shown in Fig. S3(a), which is a direct indication that entanglement can be achieved for extended time periods only at the ENZ case.

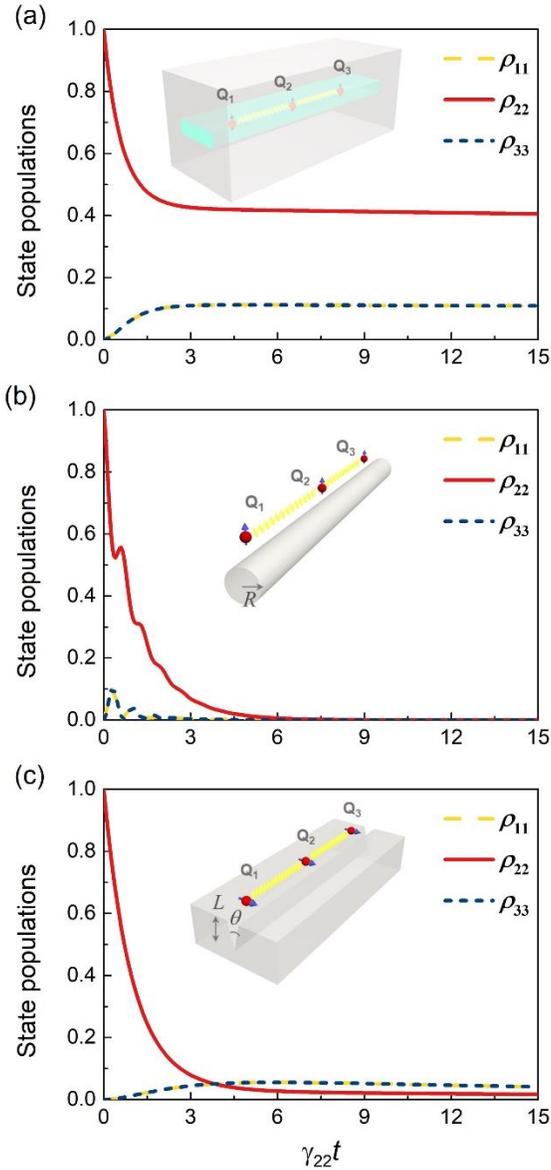

FIG. S3. Time evolution of each state population when three qubits are placed in the ENZ (a), cylindrical rod (b), and groove (c) plasmonic waveguide. The inter-qubits separation distance is $d = 200$nm, and the initially excited qubit is set to be $Q_2$ with $\rho_{22}(0) = 1$. The x-axes are normalized to $\gamma_{22}$, i.e., the computed emission decay rate of $Q_2$ in each scenario.

Finally, the four-qubit entanglement can also be characterized by the genuine multipartite entanglement ($E_G^{(2)}$) when the metric in Eq. (29) is generalized to four qubits. Assuming that Q$_2$ is initially excited, and each inter-qubit separation distance is $d = 300$ nm, we compute and plot in Fig. S4 the transient genuine multipartite entanglement $E_G^{(2)}$ for the ENZ plasmonic waveguide and compare the results to the finite groove and rod plasmonic waveguides or free space. It is clearly demonstrated that the $E_G^{(2)}$ results provide the same information with the four qubit negativity results presented in the main paper. In addition, the $E_G^{(2)}$ values are superior in the ENZ system, meaning that an enhanced quadripartite entanglement with extended time durations is realized. The corresponding time dependence of different state populations $\rho_{11}$, $\rho_{22}$, $\rho_{33}$ and $\rho_{44}$ in such ENZ-based four-qubit system is calculated and shown in Fig. S5.

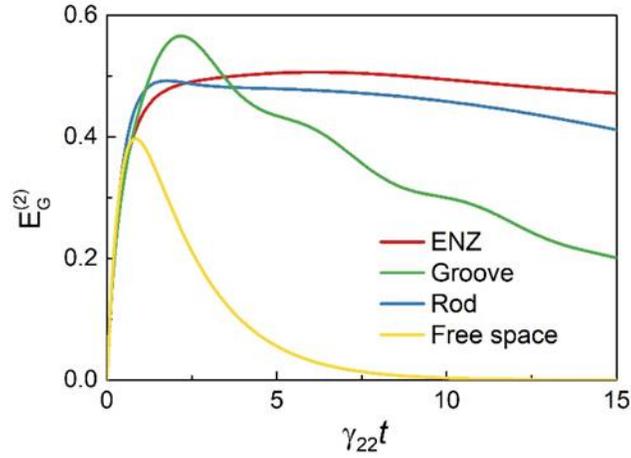

FIG. S4. Genuine multipartite entanglement $E_G^{(2)}$ of four qubits placed in ENZ, groove, and rod plasmonic waveguides, or free space. The x-axes are normalized to $\gamma_{22}$, i.e., the computed emission decay rate of Q$_2$ in each scenario.

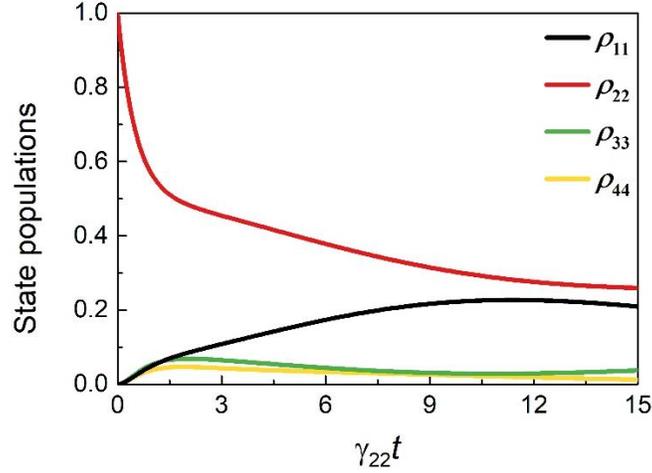

FIG. S5. Time evolution of each state population when four qubits are placed in the ENZ plasmonic waveguide. The x-axes are normalized to $\gamma_{22}$, i.e., the computed emission decay rate of Q$_2$ in each scenario.